\documentclass[3p,twocolumn]{elsarticle}

\pdfoutput=1 

\biboptions{sort&compress}

\usepackage{graphicx}
\usepackage{amssymb}
\usepackage{amsmath}
\usepackage{url}
\usepackage{cancel}
\usepackage{morefloats}
\usepackage{placeins}
\usepackage{array}
\usepackage[colorlinks,breaklinks]{hyperref}
\hypersetup{linkcolor=blue,citecolor=blue,filecolor=black,urlcolor=blue}

\makeatletter
\def\NAT@def@citea{\def\@citea{\NAT@separator}}
\makeatother

\usepackage{hhline}

\newcommand{\avgNcoll}{\mbox{$\left< N_{\mathrm{coll}}\right>$}}
\newcommand{\Ncoll}{\mbox{$N_{\mathrm{coll}}$}}

\newcommand{\Rcollcent}{\mbox{$R_{\mathrm{coll}}$}}
\newcommand{\Npart}{\mbox{$N_{\mathrm{part}}$}}
\newcommand{\TAAb}{\mbox{$T_{p\mathrm{A}}(b)$}}
\newcommand{\TAA}{\mbox{$T_{p\mathrm{A}}$}}

\newcommand{\Rcp}{\mbox{$R_{\rm CP}$}}

\newcommand{\RFour}{\mbox{$R = 0.4$}}

\newcommand{\kt}{\mbox{$k_{t}$}}

\newcommand{\pp}{\mbox{$pp$}}
\newcommand{\pPb}{\mbox{$p$+Pb}}

\newcommand{\PbPb}{\mbox{Pb+Pb}}
\newcommand{\dAu}{\mbox{$d$+Au}}

\newcommand{\sqrtsNN}{\mbox{$\sqrt{s_{\mathrm{NN}}}$}}
\newcommand{\sqrts}{\mbox{$\sqrt{s}$}}

\newcommand{\Rcoll}{\mbox{$R_{\mathrm{coll}}$}}

\newcommand{\RCP}{\mbox{$R_{\rm CP}$}}
\newcommand{\RpPb}{\mbox{$R_{p{\rm Pb}}$}}

\newcommand{\delRCP}{\mbox{$\delta R_{\rm CP}$}}
\newcommand{\delRpPb}{\mbox{$\delta R_{\rm pPb}$}}

\newcommand{\TAB}{\mbox{$T_{\mathrm{pA}}$}}
\newcommand{\xT}{\mbox{$x_{\mathrm{T}}$}}
\newcommand{\Njet}{\mbox{$N^{\mathrm{jet}}$}}
\newcommand{\Njetreco}{\mbox{$N^{\mathrm{jet}}_{\mathrm{reco}}$}}
\newcommand{\SETPb}{\mbox{$\Sigma{E}_\mathrm{T}^\mathrm{Pb}$}}

\usepackage[cernpreprint]{atlascover}

\usepackage{atlasphysics}
\usepackage{atlasunit}
\usepackage{atlasmisc}

\AtlasTitle{Centrality and rapidity dependence of inclusive jet production in $\sqrtsNN
=5.02$~\TeV\ proton--lead collisions with the ATLAS detector}

\PreprintIdNumber{CERN-PH-EP-2014-275}

\AtlasJournal{Phys. Lett. B}

\AtlasAbstract{ %
Measurements of the centrality and rapidity dependence of inclusive
jet production in $\sqrtsNN = 5.02$~\TeV\ proton--lead (\pPb)
collisions and the jet cross-section in $\sqrt{s} = 2.76$~\TeV\
proton--proton collisions are presented. These quantities are measured
in datasets corresponding to an integrated luminosity of
$27.8$~nb$^{-1}$ and $4.0$~pb$^{-1}$, respectively, recorded with the
ATLAS detector at the Large Hadron Collider in 2013. The \pPb\
collision centrality was characterised using the total transverse
energy measured in the pseudorapidity interval $-4.9 < \eta < -3.2$ in
the direction of the lead beam. Results are presented for the
double-differential per-collision yields as a function of jet rapidity
and transverse momentum (\pT) for minimum-bias and
centrality-selected \pPb\ collisions, and are compared to the jet rate
from the geometric expectation. The total jet yield in minimum-bias
events is slightly enhanced above the expectation in a \pT-dependent
manner but is consistent with the expectation within uncertainties.
The ratios of jet spectra from different centrality selections show a
strong modification of jet production at all \pT\ at forward
rapidities and for large \pT\ at mid-rapidity, which manifests as a
suppression of the jet yield in central events and an enhancement in
peripheral events. These effects imply that the factorisation between
hard and soft processes is violated at an unexpected level in
proton--nucleus collisions. Furthermore, the modifications at forward
rapidities are found to be a function of the total jet energy only,
implying that the violations may have a simple dependence on the hard
parton--parton kinematics.
 
}

\begin{document}

\title{Centrality and rapidity dependence of inclusive jet production in $\sqrtsNN
=5.02$~\TeV\ proton--lead collisions with the ATLAS detector}

\author{ATLAS Collaboration}

\begin{abstract}
Measurements of the centrality and rapidity dependence of inclusive
jet production in $\sqrtsNN = 5.02$~\TeV\ proton--lead (\pPb)
collisions and the jet cross-section in $\sqrt{s} = 2.76$~\TeV\
proton--proton collisions are presented. These quantities are measured
in datasets corresponding to an integrated luminosity of
$27.8$~nb$^{-1}$ and $4.0$~pb$^{-1}$, respectively, recorded with the
ATLAS detector at the Large Hadron Collider in 2013. The \pPb\
collision centrality was characterised using the total transverse
energy measured in the pseudorapidity interval $-4.9 < \eta < -3.2$ in
the direction of the lead beam. Results are presented for the
double-differential per-collision yields as a function of jet rapidity
and transverse momentum (\pT) for minimum-bias and
centrality-selected \pPb\ collisions, and are compared to the jet rate
from the geometric expectation. The total jet yield in minimum-bias
events is slightly enhanced above the expectation in a \pT-dependent
manner but is consistent with the expectation within uncertainties.
The ratios of jet spectra from different centrality selections show a
strong modification of jet production at all \pT\ at forward
rapidities and for large \pT\ at mid-rapidity, which manifests as a
suppression of the jet yield in central events and an enhancement in
peripheral events. These effects imply that the factorisation between
hard and soft processes is violated at an unexpected level in
proton--nucleus collisions. Furthermore, the modifications at forward
rapidities are found to be a function of the total jet energy only,
implying that the violations may have a simple dependence on the hard
parton--parton kinematics.

\end{abstract}
\maketitle

\section{Introduction}
Proton--lead (\pPb) collisions at the Large Hadron Collider (LHC)
provide an excellent opportunity to study hard scattering processes
involving a nuclear target~\cite{Salgado:2011wc}. Measurements of jet
production in \pPb\ collisions provide a valuable benchmark for
studies of jet quenching in lead--lead collisions by, for example,
constraining the impact of nuclear parton distributions on inclusive
jet yields. However, \pPb\ collisions also allow the study of possible
violations of the QCD factorisation between hard and soft processes
which may be enhanced in collisions involving nuclei.

Previous studies in deuteron--gold (\dAu) collisions at the
Relativistic Heavy Ion Collider (RHIC) observed such violations,
manifested in the suppressed production of very forward hadrons with
transverse momenta up to
4~\GeV~\cite{Arsene:2004ux,Adler:2004eh,Adams:2006uz}. Studies of
forward dihadron angular correlations at RHIC also showed a much
weaker dijet signal in \dAu\ collisions than in
\pp\ collisions~\cite{Adams:2006uz,Adare:2011sc}. These effects have
been attributed to the saturation of the parton distributions in the
gold
nucleus~\cite{JalilianMarian:2005jf,Gelis:2010nm,Kharzeev:2004yx}, to
the modification of the nuclear parton distribution function
\cite{Eskola:2009uj}, to the higher-twist contributions to the
cross-section enhanced by the forward kinematics of the measurement
\cite{Kopeliovich:2005ym}, or to the presence of a large nucleus
\cite{Qiu:2004da}. The extended kinematic reach of \pPb\ measurements
at the LHC allows the study of hard scattering processes that produce
forward hadrons or jets over a much wider rapidity and transverse
momentum range. Such measurements can determine whether the
factorisation violations observed at RHIC persist at higher energy
and, if so, how the resulting modifications vary as a function of
particle or jet momentum and rapidity. The results of such
measurements could test the competing descriptions of the RHIC results
and, more generally, provide new insight into the physics of hard
scattering processes involving a nuclear target.

This paper reports the centrality dependence of inclusive jet
production in \pPb\ collisions at a nucleon--nucleon centre-of-mass
energy $\sqrt{s_\mathrm{NN}} = 5.02$~\TeV. The measurement was
performed using a dataset corresponding to an integrated luminosity of
27.8~${\rm nb}^{-1}$ recorded in 2013. The \pPb\ jet yields were
compared to a nucleon--nucleon reference constructed from a
measurement of jet production in \pp\ collisions at a centre-of-mass
energy $\sqrt{s} = 2.76$~\TeV\ using a dataset corresponding to an
integrated luminosity of $4.0$ pb$^{-1}$ also recorded in 2013. Jets
were reconstructed from energy deposits measured in the calorimeter
using the anti-\kt\ algorithm with radius parameter
\RFour\ \cite{Cacciari:2008gp}.

The centrality of \pPb\ collisions was characterised using the total
transverse energy measured in the pseudorapidity\footnote{ATLAS uses a
  right-handed coordinate system with its origin at the nominal
  interaction point (IP) in the centre of the detector and the
  $z$-axis along the beam pipe. The $x$-axis points from the IP to the
  centre of the LHC ring, and the $y$-axis points upward. Cylindrical
  coordinates $(r,\phi)$ are used in the transverse plane, $\phi$
  being the azimuthal angle around the beam pipe.  The pseudorapidity
  is defined in laboratory coordinates in terms of the polar angle
  $\theta$ as $\eta = -\ln\tan(\theta/2)$.  During 2013 $p$+Pb
  data-taking, the beam directions were reversed approximately
  half-way through the running period, but in presenting results the
  direction of the proton beam is always chosen to point to positive
  $\eta$.  } interval $-4.9 < \eta < -3.2$ in the direction of the
lead beam. Whereas in nucleus--nucleus collisions centrality reflects
the degree of nuclear overlap between the colliding nuclei, centrality
in \pPb\ collisions is sensitive to the multiple interactions between
the proton and nucleons in the lead nucleus. Centrality has been
successfully used at lower energies in \dAu\ collisions at RHIC as an
experimental handle on the collision
geometry~\cite{Adler:2006wg,Back:2004mr,Arsene:2004ux}.

A Glauber model \cite{Miller:2007ri} was used to determine the average
number of nucleon--nucleon collisions, \avgNcoll, and the mean value
of the overlap function, $\TAAb = \int_{-\infty}^{+\infty}\rho(b,
z)\mathrm{d}z$, where $\rho(b,z)$ is the nucleon density at impact
parameter $b$ and longitudinal position $z$, in each centrality
interval. Per-event jet yields, $(1/\Nevt)
(\mathrm{d}^2N_{\mathrm{jet}}/\mathrm{d}\pt \mathrm{d}y^*)$, were
measured as a function of jet centre-of-mass rapidity,\footnote{The
  jet rapidity $y^*$ is defined as $y^* = 0.5\ln\frac{E + p_z}{E -
    p_z}$ where $E$ and $p_z$ are the energy and the component of the
  momentum along the proton beam direction in the nucleon--nucleon
  centre-of-mass frame.} $y^*$, and transverse momentum, \pT, where
$N_\mathrm{jet}$ is the number of jets measured in $\Nevt$
\pPb\ events analysed. The centrality dependence of the per-event jet
yields was evaluated using the nuclear modification factor,

\begin{equation}
\RpPb \equiv \dfrac{1}{\TAA} \dfrac{\left.\left(1/N_\mathrm{evt}\right) \;
  \mathrm{d}^2N_{\mathrm{jet}}/\mathrm{d}\pt \mathrm{d}y^*\right|_\mathrm{cent}}
{ \mathrm{d}^2\sigma_{\mathrm{jet}}^{pp}/\mathrm{d}\pt \mathrm{d}y^*},
\end{equation}

\noindent for a given centrality selection ``cent'', where
$\mathrm{d}^2\sigma_{\mathrm{jet}}^{pp}/\mathrm{d}\pt \mathrm{d}y^*$
is determined using the jet cross-section measured in \pp\ collisions
at $\sqrt{s} = 2.76$~\TeV. The factor \RpPb\ quantifies the absolute
modification of the jet rate relative to the geometric expectation. In
each centrality interval, the geometric expectation is the jet rate
that would be produced by an incoherent superposition of the number of
nucleon--nucleon collisions corresponding to the mean nuclear
thickness in the given class of \pPb\ collisions.

Results are also presented for the central-to-peripheral ratio,

\begin{equation}
R_\mathrm{CP} \equiv \dfrac{1}{\Rcoll} \dfrac{\left.\left(1/N_\mathrm{evt}\right) \; 
  \mathrm{d}^2N_{\mathrm{jet}}/\mathrm{d}\pt \mathrm{d}y^*\right|_\mathrm{cent}}
{\left.\left(1/N_\mathrm{evt}\right) \; \mathrm{d}^2N_{\mathrm{jet}}/\mathrm{d}\pt \mathrm{d}y^*\right|_\mathrm{peri}},
\end{equation}

\noindent where \Rcoll\ represents the ratio of \avgNcoll\ in a given
centrality interval to that in the most peripheral interval,
$\Rcollcent \equiv \left< N_{\mathrm{coll}}^{\mathrm{cent}} \right> /
\left< \smash{N_{\mathrm{coll}}^{\mathrm{peri}}} \right>$. The
\Rcp\ ratio is sensitive to relative deviations in the jet rate from
the geometric expectation between the \pPb\ event centralities. The
\RpPb\ and \Rcp\ measurements are presented as a function of inclusive
jet $y^*$ and \pT.

For the 2013 \pPb\ run, the LHC was configured with a 4~\TeV\ proton
beam and a 1.57~\TeV\ per-nucleon Pb beam that together produced
collisions with $\sqrtsNN = 5.02$~\TeV\ and a rapidity shift of the
centre-of-mass frame of $0.465$ units relative to the \mbox{ATLAS}
rest frame. The run was split into two periods, with the directions of
the proton and lead beams being reversed at the end of the first
period. The first period provided approximately 55\% of the integrated
luminosity with the Pb beam travelling to positive rapidity and the
proton beam to negative rapidity, and the second period provided the
remainder with the beams reversed. The analysis in this paper uses the
events from both periods of data-taking and $y^*$ is defined so that
$y^*>0$ always refers to the downstream proton direction.

\section{Experimental setup}
\label{sec:expsetup}

The measurements presented in this paper were performed using the
ATLAS inner detector (ID), calorimeters, minimum-bias trigger
scintillator (MBTS), and trigger and data acquisition
systems~\cite{Aad:2008zzm}. The ID measures charged particles within
$|\eta|<2.5$ using a combination of silicon pixel detectors, silicon
microstrip detectors, and a straw-tube transition radiation tracker,
all immersed in a 2~T axial magnetic field~\cite{Aad:2010bx}.  The
calorimeter system consists of a liquid argon (LAr) electromagnetic
(EM) calorimeter covering $|\eta|<3.2$, a steel/scintillator sampling
hadronic calorimeter covering $|\eta| < 1.7$, a LAr hadronic
calorimeter covering $1.5 < |\eta| < 3.2$, and two LAr electromagnetic
and hadronic forward calorimeters (FCal) covering $3.2 < |\eta| <
4.9$.  The EM calorimeters use lead plates as the absorbers and are
segmented longitudinally in shower depth into three compartments with
an additional presampler layer in front for $|\eta| < 1.8$. The
granularity of the EM calorimeter varies with layer and
pseudorapidity. The middle sampling layer, which typically has the
largest energy deposit in EM showers, has a $\Delta \eta \times \Delta
\phi$ granularity of $0.025 \times 0.025$ within $|\eta| < 2.5$. The
hadronic calorimeter uses steel as the absorber and has three segments
longitudinal in shower depth with cell sizes $\Delta \eta \times
\Delta \phi = 0.1 \times 0.1$ for $|\eta| < 2.5$\footnote{An exception
  is the third (outermost) sampling layer, which has a segmentation of
  $0.2 \times 0.1$ up to $|\eta| = 1.7$.}  and $0.2 \times 0.2$ for
$2.5 < |\eta| < 4.9$. The two FCal modules are composed of tungsten
and copper absorbers with LAr as the active medium, which together
provide ten interaction lengths of material.  The MBTS detects charged
particles over $2.1 < |\eta| < 3.9$ using two hodoscopes of 16
counters each, positioned at $z = \pm 3.6\,$m.

The \pPb\ and \pp\ events used in this analysis were recorded using a
combination of minimum-bias (MB) and jet triggers
\cite{Aad:2012xs}. In \pPb\ data-taking, the MB trigger required hits
in at least one counter in each side of the MBTS detector. In
\pp\ collisions the MB condition was the presence of hits in the pixel
and microstrip detectors reconstructed as a track by the high-level
trigger system. Jets were selected using high-level jet triggers
implemented with a reconstruction algorithm similar to the procedure
applied in the offline analysis. In particular, it used the
anti-\kt\ algorithm with \RFour, a background subtraction procedure,
and a calibration of the jet energy to the full hadronic scale. The
high-level jet triggers were seeded from a combination of low-level MB
and jet hardware-based triggers. Six jet triggers with transverse
energy thresholds ranging from 20~\GeV\ to 75~\GeV\ were used to
select jets within $|\eta| < 3.2$ and a separate trigger with a
threshold of 15~\GeV\ was used to select jets with $3.2 < |\eta| <
4.9$. The triggers were prescaled in a fashion which varied with time
to accommodate the evolution of the luminosity within an LHC fill.

\section{Data selection}

\label{sec:dataselection}

In the offline analysis, charged-particle tracks were reconstructed in
the ID with the same algorithm used in
\pp\ collisions~\cite{Aad:2010ac}.  The \pPb\ events used for this
analysis were required to have a reconstructed vertex containing at
least two associated tracks with $\pt > 0.1$~\GeV, at least one hit in
each of the two MBTS hodoscopes, and a difference between times
measured on the two MBTS sides of less than 10~ns. Events containing
multiple \pPb\ collisions (pileup) were suppressed by rejecting events
having two or more reconstructed vertices, each associated with
reconstructed tracks with a total transverse momentum scalar sum of at
least $5$~\GeV. The fraction of events with one $p$+Pb interaction
rejected by this requirement was less than $0.1$\%. Events with a
pseudorapidity gap (defined by the absence of clusters in the
calorimeter with more than $0.2$~\GeV\ of transverse energy) of
greater than two units on the Pb-going side of the detector were also
removed from the analysis. Such events arise primarily from
electromagnetic or diffractive excitation of the proton. After
accounting for event selection, the number of \pPb\ events sampled by
the highest-luminosity jet trigger (which was unprescaled) was 53
billion. The event selection criteria described here were designed to
select a sample of \pPb\ events to which a centrality analysis can be
applied and for which meaningful geometric parameters can be
determined.

The \pp\ events used in this analysis were required to have a
reconstructed vertex, with the same definition as the vertices in
\pPb\ events above. No other requirements were applied.

\section{Centrality determination}

\begin{figure}[!t]
\centering
\includegraphics[width=1\columnwidth]{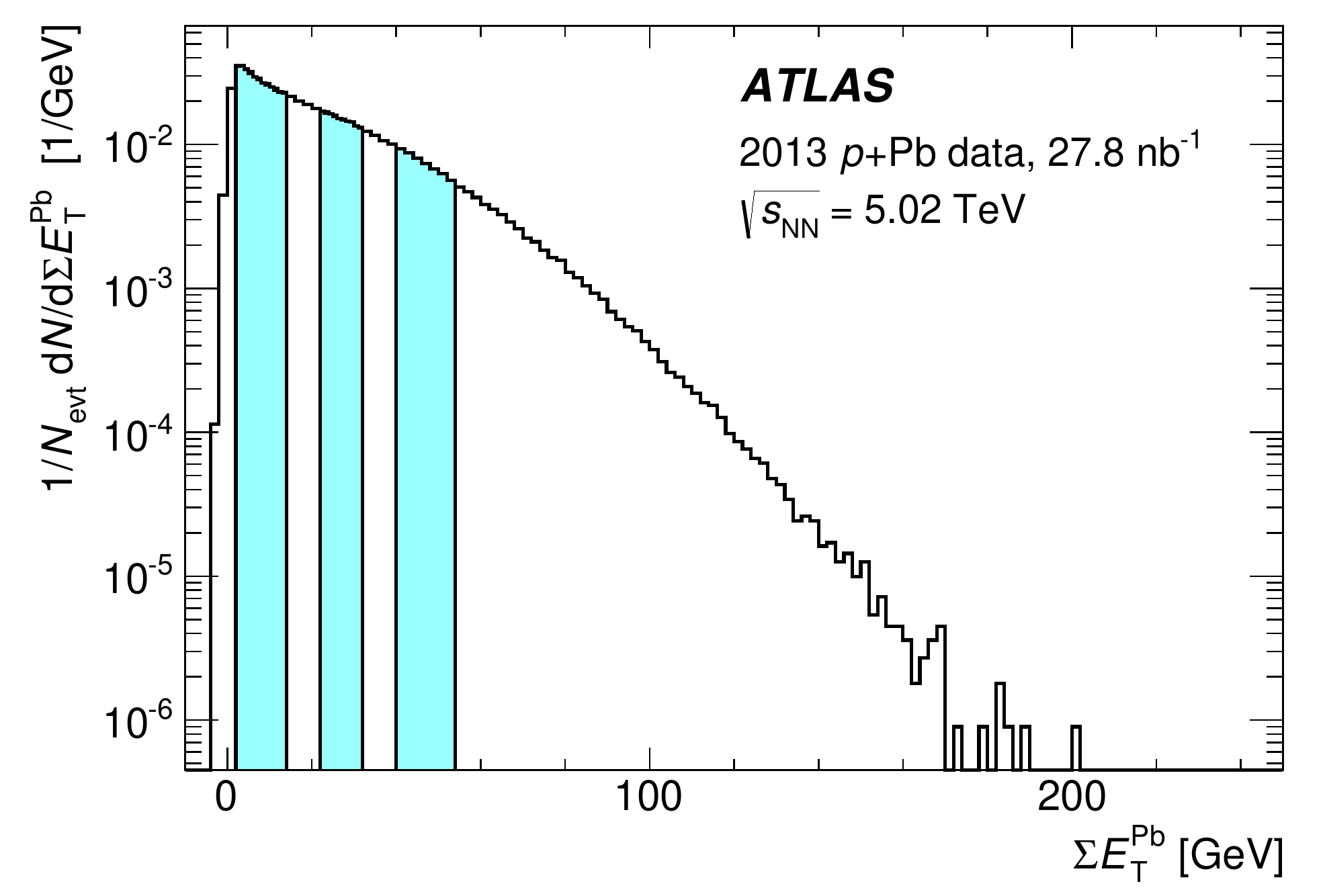}
\caption{\label{fig:fcalet} Distribution of \SETPb\ for minimum-bias
  \pPb\ collisions recorded during the 2013 run, measured in the FCal
  at $-4.9 < \eta < -3.2$ in the Pb-going direction. The vertical
  divisions correspond to the six centrality intervals used in this
  analysis. From right to left, the regions correspond to centrality
  intervals of 0--10\%, 10--20\%, 20--30\%, 30--40\%, 40--60\% and
  60--90\%.  }
\end{figure}

The centrality of the \pPb\ events selected for analysis was
characterised by the total transverse energy \SETPb\ in the FCal
module on the Pb-going side. The \SETPb\ distribution for minimum-bias
\pPb\ collisions passing the event selection described in
Section~\ref{sec:dataselection} is presented in
Fig.~\ref{fig:fcalet}. Following standard
techniques~\cite{ATLAS:2011ag}, centrality intervals were defined in
terms of percentiles of the \SETPb\ distribution after accounting for
an estimated inefficiency of $\left(2\pm 2\right)\%$ for inelastic
\pPb\ collisions to pass the applied event selection. The following
centrality intervals were used in this analysis, in order from the
most central to the most peripheral: 0--10\%, 10--20\%, 20--30\%,
30--40\%, 40--60\%, and 60--90\%, with the 60--90\% interval serving
as the reference in the \RCP\ ratio. Events with a centrality beyond
90\% were not used in the analysis, since the uncertainties on the
composition of the event sample and in the determination of the
geometric quantities are large for these events.

A Glauber Monte Carlo (MC) \cite{Miller:2007ri} analysis was used to
calculate \Rcoll\ and \TAA\ for each centrality interval. First, a
Glauber MC program~\cite{Alver:2008aq} was used to simulate the
geometry of inelastic \pPb\ collisions and calculate the probability
distribution of the number of nucleon participants \Npart,
$P(\Npart)$. The simulations used a Woods--Saxon nuclear density
distribution and an inelastic nucleon--nucleon cross-section,
$\sigma_\mathrm{NN}$, of $70 \pm 5$~mb. Separately,
PYTHIA 8~\cite{Sjostrand:2006za,Sjostrand:2007gs}
simulations of $4$~\TeV\ on $1.57$~\TeV\ \pp\ collisions provided a
detector-level \SETPb\ distribution for nucleon--nucleon collisions,
to be used as input to the Glauber model. This distribution was fit to
a gamma distribution.

Then, an extension of the wounded-nucleon (WN)~\cite{Bialas:1976ed}
model that included a non-linear dependence of \SETPb\ on \Npart\ was
used to define \Npart-dependent gamma distributions for \SETPb, with
the constraint that the distributions reduce to the PYTHIA
distribution for $\Npart = 2$. The non-linear term accounted for the
possible variation of the effective FCal acceptance resulting from an
$\Npart$-dependent backward rapidity shift of the produced soft
particles with respect to the nucleon--nucleon
frame~\cite{Steinberg:2007fg}. The gamma distributions were summed
over \Npart\ with a $P(\Npart)$ weighting to produce a hypothetical
\SETPb\ distribution. That distribution was fit to the measured
\SETPb\ distribution shown in Fig.~\ref{fig:fcalet} with the
parameters of the extended WN model allowed to vary freely.  The best
fit, which contained a significant non-linear term, successfully
described the \SETPb\ distribution in data over several orders of
magnitude.  From the results of the fit, the distribution of $\Npart$
values and the corresponding $\left<\Npart\right>$ were calculated for
each centrality interval. The resulting \Rcoll\ and \TAA\ values and
corresponding systematic uncertainties, which are described in
Section~\ref{sec:systerr}, are shown in Table~\ref{table:Rcollvalues}.

\begin{table}
\caption{\label{table:Rcollvalues} \newline
Average $\Rcoll$ and \TAA\ values
  for the centrality intervals used in this analysis along with total 
  systematic uncertainties. The \Rcoll\ values are with respect to 60--90\%
  events, where $\left<\Ncoll\right> = 2.98^{+0.21}_{-0.29}$. }
\begin{center}
{\setlength{\extrarowheight}{5pt}
\begin{tabular}{|c|c|c|}
\hhline{|===|} \hhline{}
Centrality & \Rcoll\ & \TAA\ [mb$^{-1}$] \\ \hline
 $0$--$90$\%   & -                         & $0.107^{+0.005}_{-0.003}$  \\ 
 $60$--$90$\%  & -                         & $0.043^{+0.003}_{-0.004}$  \\ 
 $40$--$60$\%  & $2.16^{+0.08}_{-0.07}$  & $0.092^{+0.004}_{-0.006}$  \\  
 $30$--$40$\% &  $3.00^{+0.21}_{-0.14}$  & $0.126^{+0.003}_{-0.004}$ \\ 
 $20$--$30$\% &  $3.48^{+0.33}_{-0.18}$  & $0.148^{+0.004}_{-0.002}$  \\ 
 $10$--$20$\% &  $4.05^{+0.49}_{-0.21}$  & $0.172^{+0.007}_{-0.003}$ \\ 
 $0$--$10$\% &  $4.89^{+0.83}_{-0.27}$  & $0.208^{+0.019}_{-0.005}$ \\ \hhline{|===|} \hhline{}
\end{tabular}
}
\end{center}
\end{table}

\section{Monte Carlo simulation}

The performance of the jet reconstruction procedure was evaluated
using a sample of 36~million events in which simulated $\sqrts =
5.02$~\TeV\ \pp\ hard-scattering events were overlaid with
minimum-bias \pPb\ events recorded during the 2013 run. Thus the
sample contains an underlying event contribution that is identical in
all respects to the data. The simulated events were generated using
PYTHIA~\cite{Sjostrand:2006za} (version 6.425, AUET2B
tune~\cite{ATL-PHYS-PUB-2012-003}, CTEQ6L1 parton distribution
functions~\cite{Pumplin:2002vw}) and the detector effects were fully
simulated using GEANT4~\cite{Agostinelli:2002hh,Aad:2010ah}. These
events were produced for different \pt\ intervals of the
generator-level (``truth'') \RFour\ jets. In total, the
generator-level spectrum spans $10 < \pt < 10^{3}$~\GeV.  Separate
sets of 18~million events each were generated for the two different
beam directions to take into account any $z$-axis asymmetries in the
detector. For each beam direction, the four-momenta of the generated
particles were longitudinally boosted by a rapidity of $\pm 0.465$ to
match the corresponding beam conditions. The events were simulated
using detector conditions appropriate to the two periods of the 2013
\pPb\ run and reconstructed using the same algorithms as were applied
to the experimental data. A separate 9-million-event sample of fully
simulated 2.76~\TeV\ PYTHIA \pp\ hard scattering events (with the same
version, tune and parton distribution function set) was used to
evaluate the jet performance in $\sqrt{s} =
2.76$~\TeV\ \pp\ collisions during 2013 data-taking.

\section{Jet reconstruction and performance}
\label{sec:performance}

The jet reconstruction and underlying event subtraction procedures
were adapted from those used by ATLAS in \PbPb\ collisions, which are
described in detail in Refs.~\cite{Aad:2012vca,Aad:2013sla}, and are
summarised here along with any substantial differences from the
referenced analyses.

An iterative procedure was used to obtain an event-by-event estimate
of the underlying event energy density while excluding contributions
from jets to that estimate. The modulation of the underlying event
energy density to account for potential elliptic flow was not included
in this analysis. Jets were reconstructed from the anti-\kt\ algorithm
with \RFour\ applied to calorimeter cells grouped into
$\Delta\eta\times\Delta\phi=0.1\times0.1$ towers, with the final jet
kinematics calculated from the background-subtracted energy in the
cells contained in the jet. The rate of jets reconstructed from the
underlying event fluctuations of soft particles was negligible in the
kinematic range studied and therefore no attempt to reject them was
made. The mean subtracted transverse energy in \pPb\ collisions was
$2.4$~\GeV\ ($1.4$~\GeV) for jets with $\left|y^*\right| < 1$ ($y^* >
3$). In \pp\ collisions, this procedure simply subtracts the
underlying event pedestal deposited in the calorimeter which can
arise, in part, from the presence of additional \pp\ interactions in
the same crossing (in-time pileup).

Following the above jet reconstruction, a small correction, typically
a few percent, was applied to the transverse momentum of those jets
which did not overlap with a region excluded from the background
determination and thus were erroneously included in the initial
estimate of the underlying event background. Then, the jet energies
were corrected to account for the calorimeter energy response using an
$\eta$- and \pT-dependent multiplicative factor that was derived from
the simulations \cite{Aad:2011he}. Following this calibration, a final
multiplicative {\em in situ} calibration was applied to account for
differences between the simulated detector response and data. The
measured \pt\ of jets recoiling against objects with an independently
calibrated energy scale -- such as $Z$ bosons, photons, or jets in a
different region of the detector -- was investigated. The {\em in
  situ} calibration, which typically differed from unity by a few
percent, was derived by comparing this \pT\ balance in \pp\ data with
that in simulations in a fashion similar to that used previously
within ATLAS~\cite{Aad:2014bia}.

The jet reconstruction performance was evaluated in the simulated
samples by applying the same subtraction and reconstruction procedure
as was applied to data. The resulting reconstructed jets with
transverse momentum $\pT^\mathrm{reco}$ were compared with their
corresponding generator jets, which were produced by applying the
anti-\kt\ algorithm to the final-state particles produced by PYTHIA,
excluding muons and neutrinos. Each generator jet was matched to a
reconstructed jet, and the \pt\ difference between the two jets was
studied as a function of the generator jet transverse momentum,
$\pT^\mathrm{gen}$, and generator jet rapidity $y^*$, and in the six
\pPb\ event centrality intervals.

The reconstruction efficiency for jets having $\pT^\mathrm{gen} >
25$~\GeV\ was found to be greater than $99$\%. The performance was
quantified by the means and standard deviations of the $\Delta \pt /
\pt \left(= \pT^\mathrm{reco}/\pT^\mathrm{gen} - 1\right)$
distributions, referred to as the jet energy scale closure and jet
energy resolution respectively. The closure in \pPb\ events was less
than $2$\% for $\pT^\mathrm{gen} > 25$~\GeV\ jets and was better than
$1$\% for $\pT^\mathrm{gen} > 100$~\GeV\ jets. At low
$\pT^\mathrm{gen}$, the energy scale closure and resolution exhibited
a weak \pPb\ centrality dependence, with differences in the closure of
up to $1$\% and differences in the resolution of up to 2\% in the most
central 0--10\% events relative to the 60--90\% peripheral events. At
high jet \pT, the response was centrality independent within
sensitivity. In \pp\ events, the closure was less than $1$\% in the
entire kinematic range studied.

In order to quantify the degree of \pt-bin migration introduced by the
detector response and reconstruction procedure, response matrices were
populated by recording the \pT\ values of each
generator--reconstructed jet pair. Separate matrices were constructed
for each $y^*$ interval and \pPb\ centrality interval used in the
analysis. The \pT\ bins used were chosen to increase with \pT\ such
that the width of each bin was $\approx0.25$ of the bin low
edge. Using this binning, the proportion of jets with reconstructed
\pT\ in the same bin as their truth \pT\ monotonically increased with
truth \pT\ and was $50$--$70$\%.

\section{Data analysis}

A combination of minimum-bias and jet triggered \pPb\ events were
selected for analysis as described in Section~\ref{sec:expsetup}. The
sampled luminosity (defined as the luminosity divided by the mean
luminosity-weighted prescale) of the jet triggers increased with
increasing \pT\ threshold. Offline jets were selected for the analysis
by requiring a match to an online jet trigger. The efficiency of the
various triggers was determined with respect to the minimum-bias
trigger and to lower threshold jet triggers. For simplicity, each
\pT\ bin used jets selected by only one jet trigger. In a given
\pT\ bin, jets were selected by the highest-threshold jet trigger for
which the efficiency was determined to be greater than $99$\% in the
bin. No additional corrections for the trigger efficiency were
applied.

The double-differential per-event jet yields in \pPb\ collisions were
constructed via

\begin{equation}
\frac{1}{\Nevt} \frac{\mathrm{d}^2\Njet}{\mathrm{d}\pT{\mathrm{d}y^{*}}} = \frac{1}{\Nevt} \frac{\Njet}{\Delta\pT\Delta{y^*}},
\end{equation}

\noindent where $\Nevt$ is the total (unprescaled) number of MB
\pPb\ events sampled, $\Njet$ is the yield of jets corrected for all
detector effects and the instantaneous trigger prescale during
data-taking, and $\Delta\pT$ and $\Delta{y^*}$ are the widths of the
\pT\ and $y^*$ bins. The centrality-dependent yields were constructed
by restricting $\Nevt$ and $\Njet$ to come from \pPb\ events within a
given centrality interval. The double-differential cross-section in
\pp\ collisions was constructed via

\begin{equation}
\frac{\mathrm{d}^{2}\sigma}{\mathrm{d}\pT{\mathrm{d}y^{*}}} = \frac{1}{L_\mathrm{int}}
\frac{\Njet}{\Delta\pT\Delta{y^*}},
\end{equation}

\noindent where $L_\mathrm{int}$ is the total integrated luminosity of
the jet trigger used in the given \pT\ bin. The \pT\ binning in the
\pp\ cross-section was chosen such that the $\xT~=~2\pT/\sqrt{s}$
binning between the \pPb\ and \pp\ datasets is the same.

Both the per-event yields in \pPb\ collisions and the cross-section in
\pp\ collisions were restricted to the \pT\ range where the MC studies
described in Section~\ref{sec:performance} show that the efficiency
for a truth jet to remain in the same \pT\ bin is $\ge 50$\%. This
\pT\ range was rapidity dependent, with the lowest \pT\ bin edge used
ranging from $50$~GeV in the most backward rapidity intervals studied
to $25$~GeV in the most forward intervals.

The measured \pPb\ and \pp\ yields were corrected for jet energy
resolution and residual distortions of the jet energy scale which
result in \pT-bin migration. For each rapidity interval, the yield was
corrected by the use of \pT-dependent (and, in the \pPb\ case,
centrality-dependent) bin-by-bin correction factors $C(\pT, y^*)$
obtained from the ratio of the reconstructed to the truth jet
\pT\ distributions for jets originating in a true $y^*$ bin, according
to

\begin{equation}
C(\pT, y^*) = \frac{N^\mathrm{jet}_\mathrm{truth}(\pT, y^*)}{N^\mathrm{jet}_\mathrm{reco}(\pT, y^*)},
\end{equation}

\noindent where $N^\mathrm{jet}_\mathrm{truth}$
($N^\mathrm{jet}_\mathrm{reco}$) is the number of truth jets in the
given $p_\mathrm{T}^\mathrm{truth}$ ($p_\mathrm{T}^\mathrm{reco}$) bin
in the corresponding MC samples.

Since the determination of the correction factors $C(\pT, y^*)$ is
sensitive to the shape of the jet spectrum in the MC sample, the
response matrices used to generate them were reweighted to provide a
better match between the reconstructed distributions in data and
simulated events. The spectrum of generator jets was weighted
jet-by-jet by the ratio of the reconstructed spectrum in data to that
in simulation. This ratio was found to be approximately linear in the
logarithm of reconstructed \pt. A separate reweighting was performed
for the \pPb\ jet yield in each centrality interval, resulting in
changes of $\le10$\% from the original correction factors before
reweighting. The resulting corrections to the \pPb\ and \pp\ yields
were at most 30\%, and were typically $\le 10$\% for jets with $\pT >
100$~\GeV.  These corrections were applied to the detector-level yield
\Njetreco\ to give the particle-level yield via

\begin{eqnarray}
\Njet = C(\pT, y^*) \Njetreco.
\end{eqnarray}

A $\sqrt{s} = 5.02$~\TeV\ \pp\ reference jet cross-section was
constructed through the use of the corrected
$2.76$~\TeV\ \pp\ cross-section and a previous ATLAS measurement of
the \xT-scaling between the inclusive jet cross-sections at $\sqrt{s}
= 2.76$~\TeV\ (measured using $0.20$~pb$^{-1}$ of data collected in
2011) and $7$~\TeV\ (measured using $37$~pb$^{-1}$ of data collected
in 2010)~\cite{Aad:2013lpa}. In this previous analysis, the
$\sqrt{s}$-scaled ratio $\rho$ of the $2.76$~\TeV\ cross-section to
that at $7$~\TeV\ was evaluated at fixed $\xT$,

\begin{equation}
\rho(\xT; y^*) = 
\left(\frac{2.76\ \mathrm{TeV}}{7\ \mathrm{TeV}}\right)^3 
\frac{\mathrm{d}^2\sigma^{2.76\ \mathrm{TeV}}/\mathrm{d}\pT{\mathrm{d}y^*}}
{\mathrm{d}^2\sigma^{7\ \mathrm{TeV}}/\mathrm{d}\pT{\mathrm{d}y^*}}
\label{equation:pp_xsect_rho_definition},
\end{equation}

\noindent where
$\mathrm{d}^2\sigma^{\sqrt{s}}/\mathrm{d}\pT{\mathrm{d}y^*}$ is the
\pp\ jet cross-section at the given centre-of-mass energy $\sqrt{s}$,
and the numerator and denominator are each evaluated at the same
\xT\ (but different $\pT =
\xT\sqrt{s}/2$). Equation~(\ref{equation:pp_xsect_rho_definition}) can
be rearranged to define the cross-section at $\sqrt{s} = 7$~\TeV\ in
terms of that at $2.76$~\TeV\ times a multiplicative factor and
divided by $\rho$.

The $\sqrt{s} = 5.02$~\TeV\ \pp\ cross-section at each \pT\ and $y^*$
value was constructed by scaling the corrected $\sqrt{s} =
2.76$~\TeV\ \pp\ cross-section measured at the equivalent
\xT\ according to

\begin{align}
& \frac{\mathrm{d}^2\sigma^{5.02\ \mathrm{TeV}}}{\mathrm{d}\pT{\mathrm{d}y^*}} \nonumber \\
&  =  \rho(\xT; y^*)^{-0.643} 
\left(\frac{2.76\ \mathrm{TeV}}{5.02\ \mathrm{TeV}}\right)^{3} 
\frac{\mathrm{d}^2\sigma^{2.76\ \mathrm{TeV}}}{\mathrm{d}\pT{\mathrm{d}y^*}} 
\label{equation:pp_xsect_5p02TeV_construction},
\end{align}

\noindent where the power $-\ln(2.76/5.02) / \ln(2.76/7) \approx
-0.643$ interpolates between $2.76$~\TeV\ and $7$~\TeV\ to
$5.02$~\TeV\ using a power-law collision energy dependence at each
\pT\ and $y^*$. Since the jet energy scale and \xT-interpolation
uncertainties are large for the \pp\ data at large rapidities ($|y^*|
> 2.8$), a $\sqrt{s} = 5.02$~\TeV\ reference is not constructed in
that rapidity region.

The \pp\ jet cross-section at $\sqrt{s} = 2.76$~\TeV\ measured with
the 2013 data was found to agree with the previous ATLAS measurement
of the same quantity~\cite{Aad:2013lpa} within the systematic
uncertainties.

\section{Systematic uncertainties}
\label{sec:systerr}

The \RCP\ and \RpPb\ measurements are subject to systematic
uncertainties arising from a number of sources: the jet energy scale
and resolution, differences in the spectral shape between data and
simulation affecting the bin-by-bin correction factors, residual
inefficiency in the trigger selection, and the estimates of the
geometric quantities \Rcoll\ (in \RCP) and \TAB\ (in \RpPb). In
addition to these sources of uncertainty, which are common to the
\RCP\ and \RpPb\ measurements, \RpPb\ is also subject to uncertainties
from the \xT-interpolation of the
$\sqrt{s}~=~2.76$~\TeV\ \pp\ cross-section to the $\sqrt{s} =
5.02$~\TeV\ centre-of-mass energy and from the integrated luminosity
of the \pp\ dataset.

Uncertainties in the jet energy scale and resolution influence the
correction of the \pPb\ and \pp\ jet spectra. The uncertainty in the
scale was taken from studies of the {\em in situ} calorimeter response
and systematic variations of the jet response in
simulation~\cite{Aad:2011he}, as well as studies of the relative
energy scale difference between the jet reconstruction procedure in
heavy-ion collisions and the procedure used by ATLAS for inclusive jet
measurements in 2.76~\TeV\ and
7~\TeV\ \pp\ collisions~\cite{Aad:2013lpa,Aad:2011fc}. The total energy scale
uncertainty in the measured \pT\ range was $\lesssim 4$\% for jets in
$\left|y^*\right| < 2.8$, and $\lesssim 7$\% for jets in
$\left|y^*\right| > 2.8$. The sensitivity of the results to the
uncertainty in the energy scale was evaluated separately for ten distinct
sources of uncertainty. Each source was treated as fully uncorrelated
with any other source, but fully correlated with itself in \pT, $\eta$,
and $\sqrt{s}$. The uncertainty in the resolution was taken from {\em in
  situ} studies of the dijet energy balance~\cite{Aad:2012ag}. The resolution
uncertainty was generally $<10$\%, except for low-\pT\ jets where it
was $<20$\%. The effects on the \RCP\ and \RpPb\ measurements were
evaluated through an additional smearing of the energy of
reconstructed jets in the simulation such that the resolution uncertainty was
added to the original resolution in quadrature.

The resulting systematic uncertainties on \RCP\ (\delRCP) and
\RpPb\ (\delRpPb) were evaluated by producing new response matrices in
accordance with each source of the energy scale uncertainty and the resolution
uncertainty, generating new correction factors, and calculating the
new \RCP\ and \RpPb\ results. Each energy scale and resolution variation was applied
to all rapidity bins and to both the \pPb\ and \pp\ response matrices
simultaneously. The uncertainty on \RCP\ and \RpPb\ from the total energy scale
uncertainty was determined by adding the effects of the ten energy scale
uncertainty sources in quadrature. Since the correction factors for
the \pPb\ spectra in different centrality intervals were affected to a
similar degree by variations in the energy scale and resolution, the effects tended to
cancel in the \RCP\ ratio, and the resulting \delRCP\ were small. The
resulting \delRpPb\ values were somewhat larger than the
\delRCP\ values due to the relative centre-of-mass shift between the
\pPb\ and \pp\ collision systems. The centrality dependence of the energy scale
and resolution uncertainties in \pPb\ events was negligible.

To achieve better correspondence with the data, the simulated jet
spectrum was reweighted to match the spectral shape in data before
deriving the bin-by-bin correction factors as described above. To
determine the sensitivity of the results to this reweighting
procedure, the slope of the fit to the ratio of the detector-level
spectrum in data to that in simulation was varied by the fit
uncertainty, and the correction factors were recomputed with this
alternative weighting. The resulting \delRpPb\ and \delRCP\ from the
nominal values were included in the total systematic uncertainty.

As the jet triggers used for the data selection were evaluated to have
greater than $99$\% efficiency in the \pT\ regions where they are used
to select jets, an uncertainty of 1\% was chosen for the centrality
selected \pPb\ yields and the \pp\ cross-section in the range $20 <
\pT < 125$~\GeV. This uncertainty was taken to be uncorrelated between
the centrality-selected \pPb\ yields and the \pp\ cross-section,
resulting in a 1.4\% uncertainty on the \RCP\ and \RpPb\ measurements.

The geometric quantities \Rcoll\ and \TAB\ and their uncertainties are
listed in Table~\ref{table:Rcollvalues}. These uncertainties arise
from uncertainties in the geometric modelling of \pPb\ collisions and
in modelling the $\Npart$ dependence of the forward particle
production measured by \SETPb. In general, the uncertainties were
asymmetric. Uncertainties in \Rcoll\ were largest for the ratio of the
most central to the most peripheral interval (0--10\%/60--90\%), where
they were $+17/-6$\%, and smallest in the 40--60\%/60--90\% ratio,
where they were $+4/-3$\%. Uncertainties in \TAB\ were largest in the
most central (0--10\%) and most peripheral (60--90\%) centrality
intervals, where the upper or lower uncertainty was as high as $10$\%,
and smaller for intervals in the middle of the \pPb\ centrality range,
where they reached a minimum of $+3/-2$\% for the 20--30\% interval.

The \xT-interpolation of the $\sqrt{s} = 2.76$~\TeV\ \pp\ jet
cross-section to $5.02$~\TeV\ is sensitive to uncertainties in
$\rho(\xT, y^*)$, the $\sqrt{s}$-scaled ratio of jet spectra at 2.76
and 7~\TeV. Following
Eq.~(\ref{equation:pp_xsect_5p02TeV_construction}), the uncertainty in
the interpolated \pp\ cross-section ($\delta\sigma^\mathrm{5.02~TeV}$)
at fixed \xT\ is related to the uncertainty in $\rho$ $(\delta\rho)$
via $(\delta\sigma^\mathrm{5.02~TeV}/\sigma^\mathrm{5.02~TeV}) = 0.643
(\delta\rho / \rho)$, where $\delta\rho$ was taken from
Ref.~\cite{Aad:2013lpa}. The values of $\delta\rho$ ranged from 5\% to
23\% in the region of the measurement and were generally larger at
lower \xT\ and at larger rapidities.

The integrated luminosity for the 2013 \pp\ dataset was determined by
measuring the interaction rate with several ATLAS subdetectors. The
absolute calibration was derived from three van der Meer
scans~\cite{vanderMeer:296752} performed during the \pp\ data-taking
in 2013 in a fashion similar to that used previously within
ATLAS~\cite{Aad:2013ucp} for \pp\ data-taking at higher energies. The
systematic uncertainty on the integrated luminosity was estimated to
be ${3.1}$\%.

The uncertainties from the jet energy scale, jet energy resolution, reweighting and \xT-interpolation
are \pT\ and $y^*$ dependent, while the uncertainties from the
trigger, luminosity, and geometric factors are not. The total
systematic uncertainty on the \RpPb\ measurement ranges from 7\% at
mid-rapidity and high \pT\ to 18\% at forward rapidities and low
\pT. In most \pT\ and rapidity bins, the dominant systematic
uncertainty on \RpPb\ is from the \xT-interpolation. The \pT- and
$y^*$- dependent systematic uncertainties on \RCP\ are small. Near
mid-rapidity or at high \pT, they are 2\%, rising to approximately
12\% at low \pT\ in forward rapidities. Thus, in most of the kinematic
region studied, the dominant uncertainty on \RCP\ is from the
geometric factors \Rcoll.

\section{Results}

\begin{figure}[!t]
\centering
\includegraphics[width=\columnwidth]{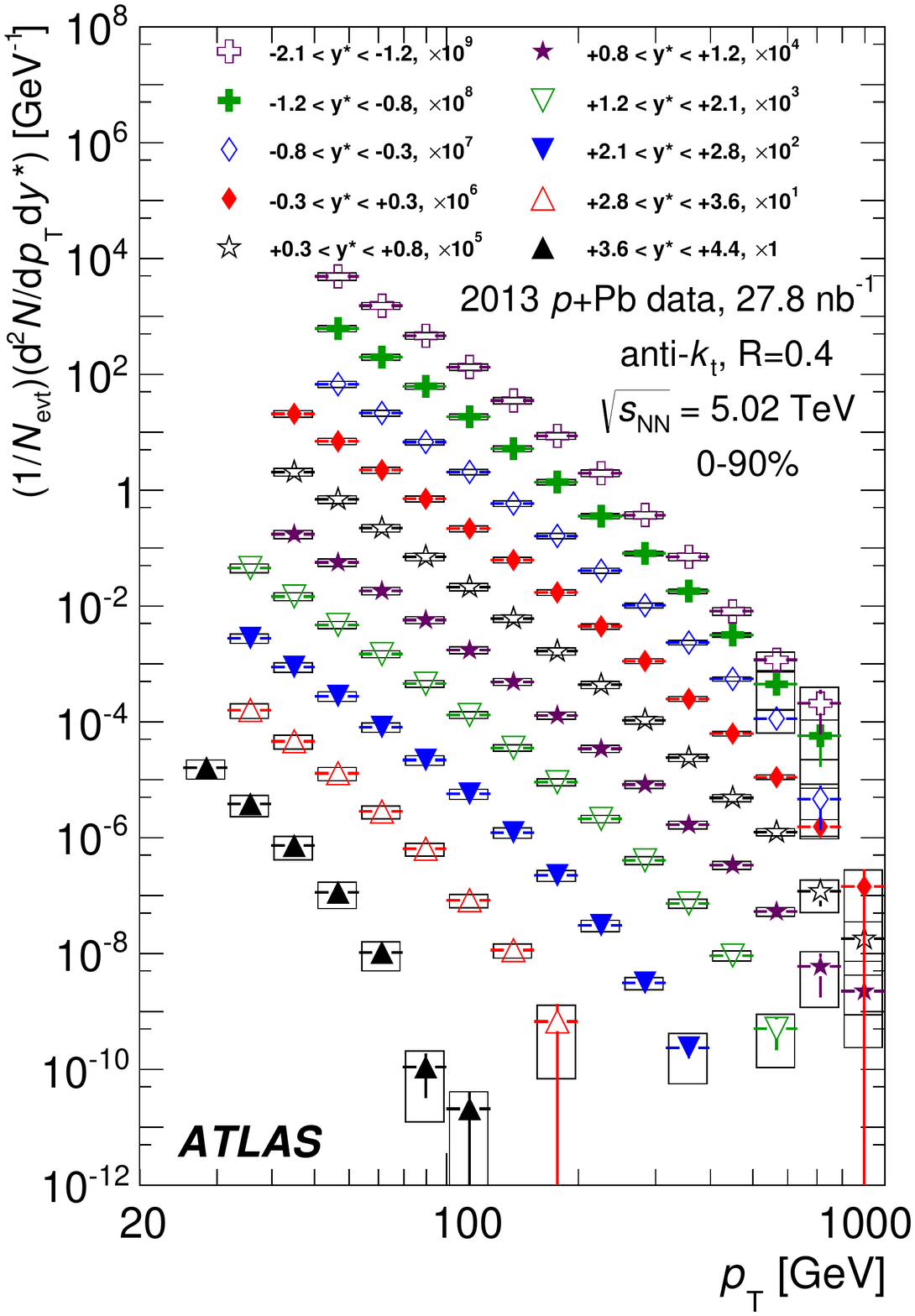}
\caption{\label{fig:pPbJetSpectrum_090} Inclusive double-differential
  per-event jet yield in 0--90\% \pPb\ collisions as a function of jet
  \pT\ in different $y^*$ bins. The yields are corrected for all
  detector effects. Vertical error bars represent the statistical
  uncertainty while the boxes represent the systematic uncertainties.}
\end{figure}

Fig.~\ref{fig:pPbJetSpectrum_090} presents the fully corrected
per-event jet yield as a function of \pt\ in 0--90\% \pPb\ collisions,
for each of the jet centre-of-mass rapidity ranges used in this
analysis. At mid-rapidity, the yields span over eight orders of
magnitude.

\begin{figure}[!t]
\centering
\includegraphics[width=1\columnwidth]{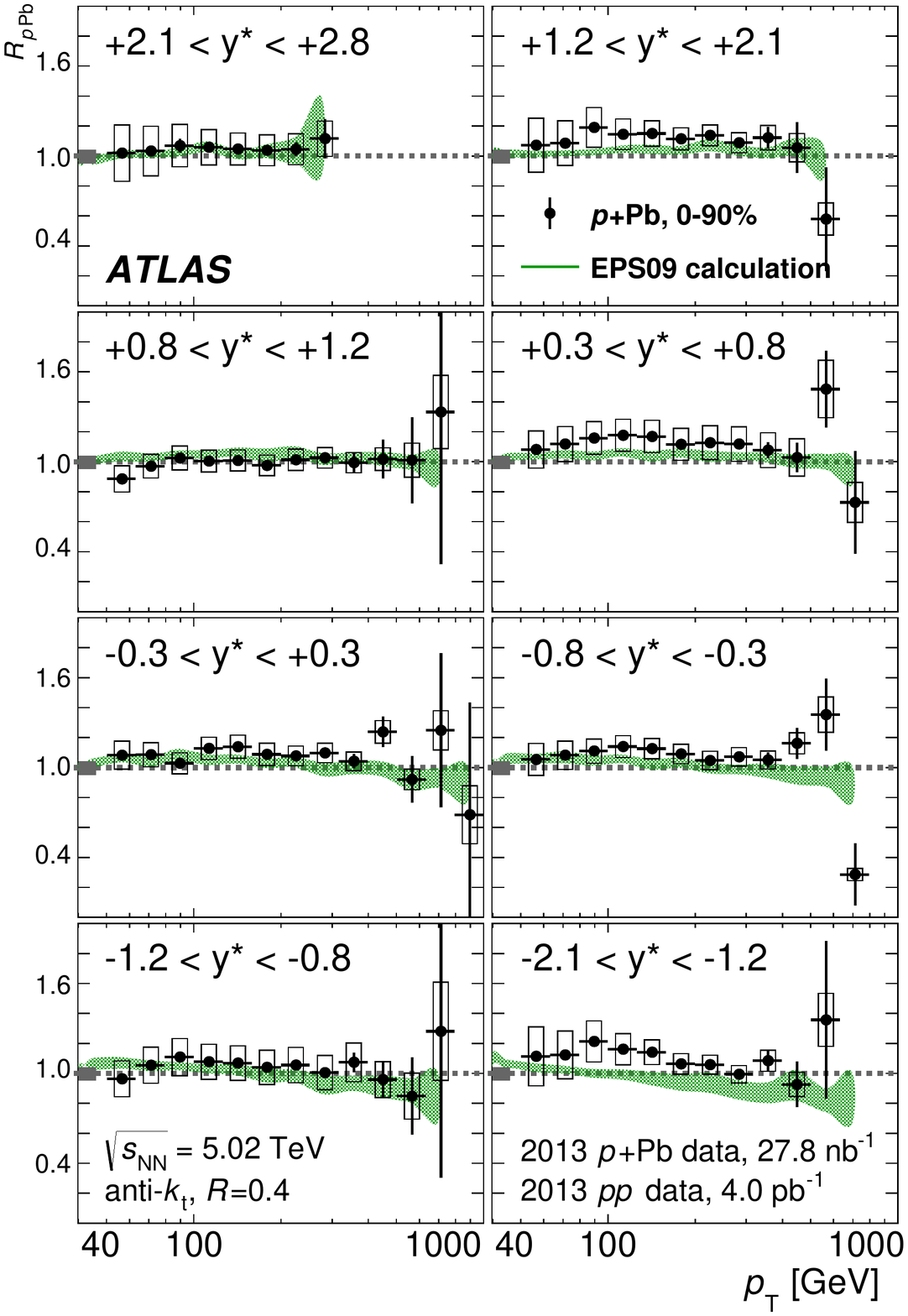}
\caption{\label{fig:pPbJetRpPb_090} Measured \RpPb\ values for $R=0.4$
  jets in 0--90\% \pPb\ collisions. Each panel shows the jet \RpPb\ in
  a different rapidity range. Vertical error bars represent the
  statistical uncertainty while the boxes represent the systematic
  uncertainties on the jet yields. The shaded box at the left edge of
  the $\RpPb = 1$ horizontal line indicates the systematic uncertainty
  on \TAA\ and the \pp\ luminosity in quadrature. The shaded band
  represents a calculation using the EPS09 nuclear parton distribution
  function set.}
\end{figure}

The jet nuclear modification factor \RpPb\ for 0--90\% \pPb\ events is
presented in Fig.~\ref{fig:pPbJetRpPb_090} in the eight rapidity bins
for which the \pp\ reference was constructed. At most rapidities
studied, the \RpPb\ values show a slight ($\approx10\%$) enhancement
above one, although many bins are consistent with unity within the
systematic uncertainties. At mid-rapidity, the \RpPb\ values reach a
maximum near $100$~\GeV.  No large modification of the total yield of
jets relative to the geometric expectation (under which $\RpPb = 1$)
is observed. The data in Fig.~\ref{fig:pPbJetRpPb_090} are compared to
a next-to-leading order perturbative QCD calculation of \RpPb\ with
the EPS09 parameterisation of nuclear parton distribution
functions~\cite{Eskola:2009uj}, using CT10~\cite{Lai:2010vv} for the
free proton parton distribution functions and following the procedure
for calculating jet production rates in \pPb\ collisions described in
Refs.~\cite{Salgado:2011wc,Albacete:2013ei}. The data are slightly
higher than the calculation, but generally compatible with it within
systematic uncertainties.



\begin{figure}[!t]
\centering
\includegraphics[width=1\columnwidth]{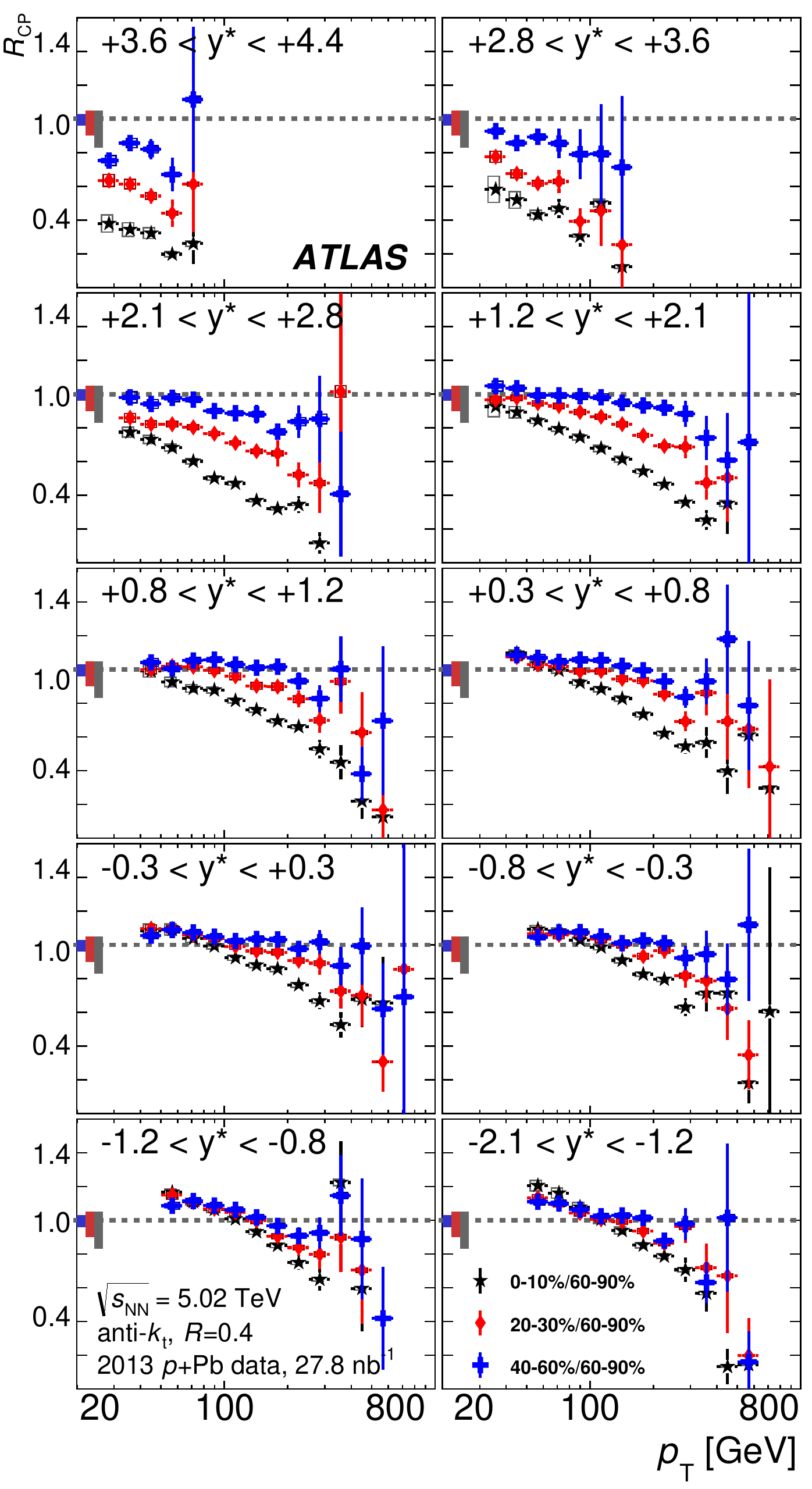}
\caption{\label{fig:pPbJetRCP} Measured \RCP\ values for $R=0.4$ jets
  in \pPb\ collisions in central (stars), mid-central (diamonds) and
  mid-peripheral (crosses) events. Each panel shows the jet \RCP\ in a
  different rapidity range. Vertical error bars represent the
  statistical uncertainty while the boxes represent the systematic
  uncertainties on the jet yields. The shaded boxes at the left edge
  of the $\RCP = 1$ horizontal line indicate the systematic
  uncertainty on \Rcoll\ for (from left to right) peripheral,
  mid-central and central events.}
\end{figure}

The central-to-peripheral ratio \RCP\ for jets in \pPb\ collisions is
summarised in Fig.~\ref{fig:pPbJetRCP}, where the \RCP\ values for
three centrality intervals are shown in all rapidity ranges
studied. The \RCP\ ratio shows a strong variation with centrality
relative to the geometric expectation, under which $\RCP = 1$. The jet
\RCP\ for 0--10\%/60--90\% events is smaller than one at all
rapidities for jet $\pT > 100$~\GeV\ and at all \pT\ at sufficiently
forward (proton-going, $y^* > 0$) rapidities. Near mid-rapidity, the
40--60\%/60--90\% \RCP\ values are consistent with unity up to
$100$--$200$~\GeV, but indicate a small suppression at higher \pT. In
all rapidity intervals studied, \RCP\ decreases with increasing
\pT\ and in increasingly more central collisions. Furthermore, at
fixed \pT, \RCP\ decreases systematically at more forward
rapidities. At the highest \pT\ in the most forward rapidity bin, the
0--10\%/60--90\% \RCP\ value is $\approx0.2$. In the backward rapidity
direction (lead-going, $y^* < 0$), \RCP\ is found to be enhanced by
$10$--$20$\% for low-\pT\ jets.
                         
\begin{figure}[!th]
\centering
\includegraphics[width=1\columnwidth]{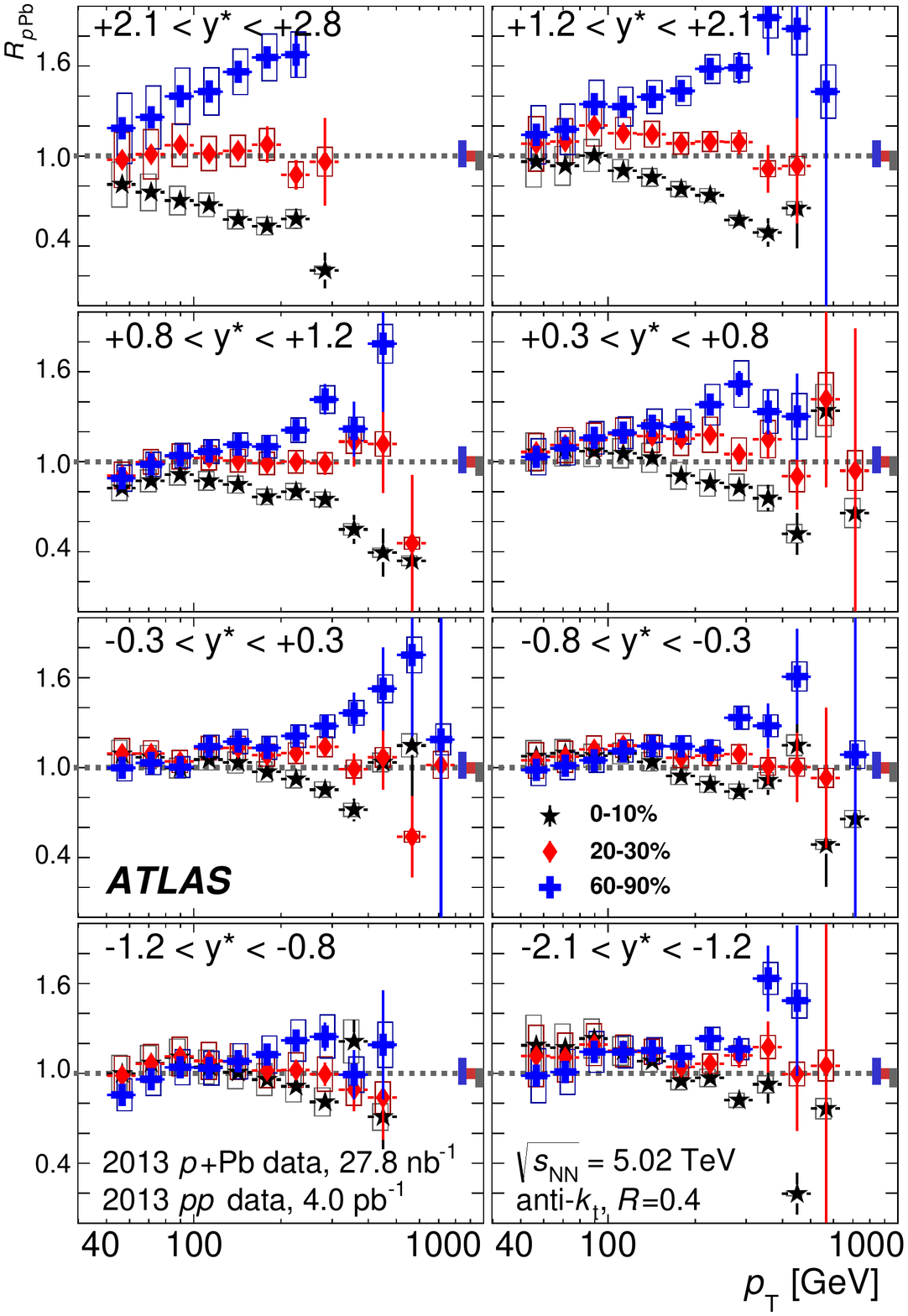}
\caption{\label{fig:pPbJetRpPb_Cent} Measured \RpPb\ values for
  $R=0.4$ jets in \pPb\ collisions in central (stars), mid-central
  (diamonds) and peripheral (crosses) events. Each panel shows the jet
  \RpPb\ in a different rapidity range.  Vertical error bars represent
  the statistical uncertainty while the boxes represent the systematic
  uncertainties on the jet yields.  The shaded boxes at the right edge
  of the $\RpPb = 1$ horizontal line indicate the systematic
  uncertainties on \TAA\ and the \pp\ luminosity added in quadrature
  for (from left to right) peripheral, mid-central and central
  events.}
\end{figure}

Fig.~\ref{fig:pPbJetRpPb_Cent} summarises the jet \RpPb\ in central,
mid-central and peripheral events in all rapidity intervals
studied. The patterns observed in the centrality-dependent
\RpPb\ values are a consequence of the near-geometric scaling of the
minimum-bias \RpPb\ values along with the strong modifications of the
central-to-peripheral ratio \RCP. At sufficiently high \pT, \RpPb\ in
central events is found to be suppressed ($\RpPb < 1$) and in
peripheral events to be enhanced ($\RpPb > 1$). Generally, these
respective deviations from the geometric expectation (under which
$\RpPb = 1$ for all centrality intervals) increase with \pT\ and, at
fixed \pT, increase as the rapidity becomes more forward. Thus, the
large effects in \RCP\ are consistent with a combination of
modifications that have opposite sign in the centrality-dependent
\RpPb\ values but have little effect on the centrality-inclusive
(0--90\%) \RpPb\ values. At backward-going rapidities ($y^* < 0$) the
\RpPb\ value for low-\pT\ jets in all centrality intervals is
consistent with unity within the uncertainties.

\begin{figure*}[!t]
\centering
\includegraphics[width=1.95\columnwidth]{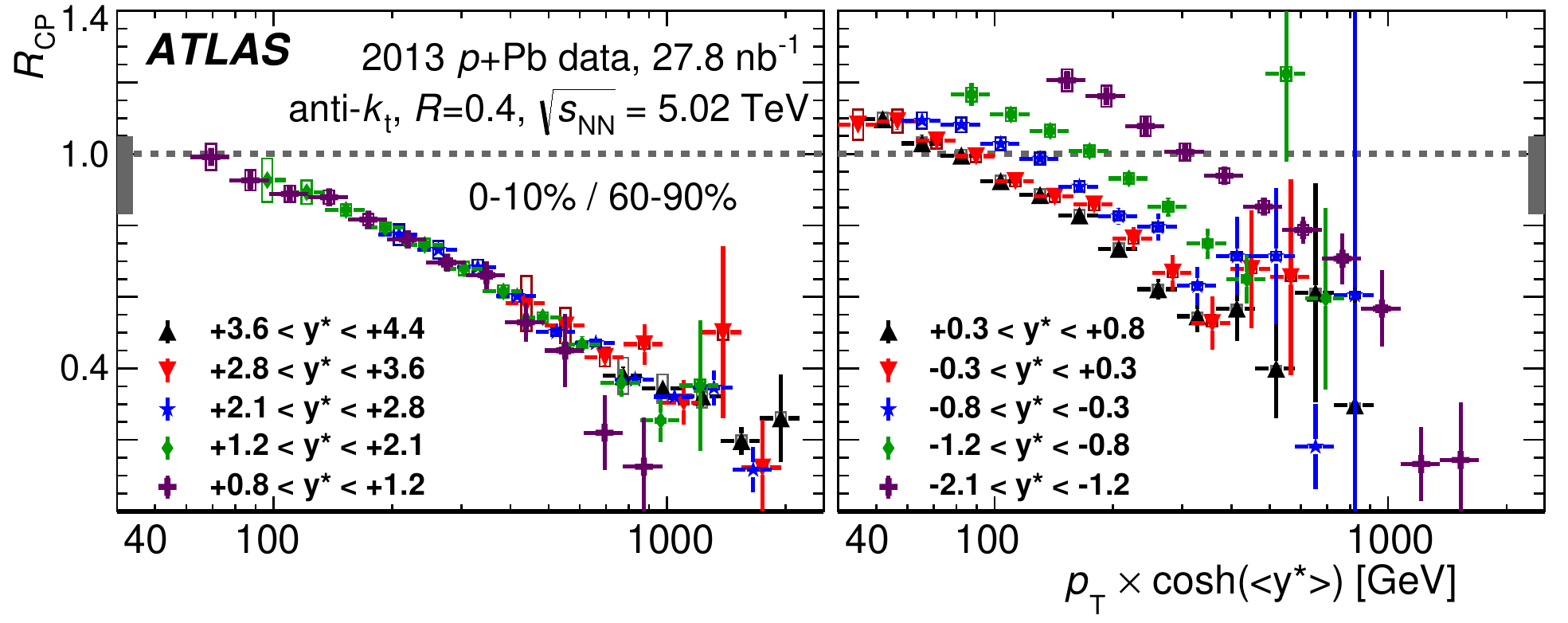}
\caption{\label{fig:pPbJetRCP_Scaling} Measured \RCP\ values for
  $R=0.4$ jets in 0--10\% \pPb\ collisions.  The panel on the left
  shows the five rapidity ranges that are the most forward-going,
  while the panel on the right shows the remaining five. The
  \RCP\ values at each rapidity are plotted as a function of
  $\pT\times\cosh(\left<y^*\right>)$, where $\left<y^*\right>$ is the
  midpoint of the rapidity bin. Vertical error bars represent the
  statistical uncertainty while the boxes represent the systematic
  uncertainties on the jet yields.  The shaded box at the left edge
  (in the left panel) and right edge (in the right panel) of the $\RCP
  = 1$ horizontal line indicates the systematic uncertainty on
  \Rcoll.}
  \end{figure*}

\begin{figure*}[!t]
\centering
\includegraphics[width=1.95\columnwidth]{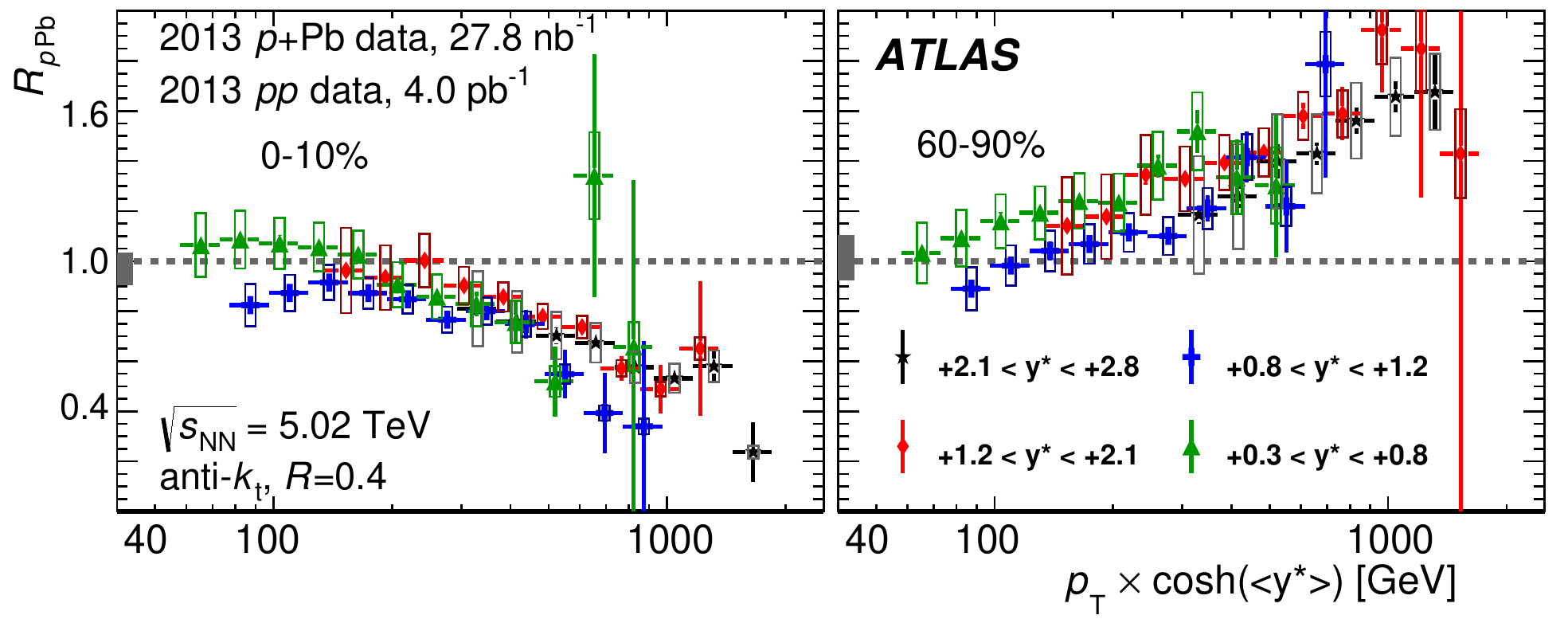}
\caption{\label{fig:pPbJetRpPb_Scaling} Measured \RpPb\ values for
  $R=0.4$ jets in \pPb\ collisions displayed for multiple rapidity
  ranges, showing 0--10\% events in the left panel and 60--90\% events
  in the right panel.  The \RpPb\ at each rapidity is plotted as a
  function of $\pT\times\cosh(\left<y^*\right>)$, where
  $\left<y^*\right>$ is the midpoint of the rapidity bin.  Vertical
  error bars represent the statistical uncertainty while the boxes
  represent the systematic uncertainties on the jet yields. The shaded
  box at the left edge of the $\RpPb = 1$ horizontal line indicates
  the systematic uncertainties on \TAB\ and the \pp\ luminosity added
  in quadrature. }
\end{figure*}

Given the observed suppression pattern as a function of jet rapidity,
in which the suppression in \RCP\ at fixed \pT\ systematically
increases at more forward-going rapidities, it is natural to ask if it
is possible to find a single relationship between the \RCP\ values in
the different rapidity intervals which is a function of jet kinematics
alone. To test this, the \RCP\ values in each rapidity bin were
plotted against the quantity $\pT\times\cosh(\left<y^*\right>) \approx
E$, where $\left<y^*\right>$ is the centre of the rapidity bin and $E$
is the total energy of the jet. In relativistic kinematics, the total
energy of a particle is given by $E = m_\mathrm{T} \cosh(y^*)$, where
the transverse mass $m_\mathrm{T} = \sqrt{m^2 + \pT^2}$. In the
kinematic range studied, the mass of the typical jet is sufficiently
small relative to its transverse momentum that approximating the
transverse mass, $m_\mathrm{T}$, with the \pT\ is reasonable. The
0--10\%/60--90\% \RCP\ versus $\pT\times\cosh(\left<y^*\right>)$ is
shown for all ten rapidity ranges in
Fig.~\ref{fig:pPbJetRCP_Scaling}. When plotted against this variable,
the \RCP\ values in each of the five forward-going rapidities ($y^* >
+0.8$) fall along the same curve, which is approximately linear in the
logarithm of $E$. This trend is also observed in the two most forward
of the remaining rapidity intervals ($-0.3 < y^* < +0.8$), but the
\RCP\ values at backward rapidities ($y^* < -0.3$) do not follow this
trend. This pattern is also observed in other centrality intervals,
albeit with a different slope in $\ln(E)$ for each centrality
interval.

These patterns suggest that the observed modifications may depend on
the initial parton kinematics, such as the longitudinal momentum
fraction of the parton originating in the proton, $x_p$. In
particular, a dependence on $x_p$ would explain why the data follow a
consistent trend vs. $\pT\times\cosh(\left<y^*\right>)$ at forward
rapidities (where jet production at a given jet energy $E$ is
dominated by $x_p \sim E/(\sqrt{s}/2)$ partons in the proton) but do
not do so at backward rapidities (where the longitudinal momentum
fraction of the parton originating in the lead nucleus,
$x_\mathrm{Pb}$, as well as $x_p$ are both needed to relate the jet
and parton kinematics).

By analogy with Fig.~\ref{fig:pPbJetRCP_Scaling} where the
\RCP\ values are plotted versus $\pT\times\cosh(\left<y^*\right>)$,
the \RpPb\ values in the four most forward-going bins studied are
plotted against this variable in
Fig.~\ref{fig:pPbJetRpPb_Scaling}. The \RpPb\ values in central and
peripheral events are shown separately. Although the systematic
uncertainties are larger on \RpPb\ than on \RCP, the observed
behaviour for jets with $\pt > 150$~\GeV\ is consistent with the
nuclear modifications depending only on the approximate total jet
energy $\pT\times\cosh(\left<y^*\right>)$. In central (peripheral)
events, the \RpPb\ values at forward rapidities are consistent with a
rapidity-independent decreasing (increasing) function of
$\pT\times\cosh(\left<y^*\right>)$. Thus, the single trend in
\RCP\ versus $\pT\times\cosh(\left<y^*\right>)$ at forward rapidities
appears to arise from opposite trends in the central and peripheral
\RpPb, both a single function of $\pT\times\cosh(\left<y^*\right>)$.

The results presented here use the standard Glauber model with fixed
$\sigma_\mathrm{NN}$ to estimate the geometric quantities. The impact
of geometric models which incorporate event-by-event changes in the
configuration of the proton wavefunction~\cite{Alvioli:2013vk} has
also been studied. Using the so called Glauber--Gribov Colour
Fluctuation model to determine the geometric parameters amplifies the
effects seen with the Glauber model. In this model, the suppression in
central events and the enhancement in peripheral events would be
increased.

\section{Conclusions}
This paper presents the results of a measurement of the centrality
dependence of jet production in \pPb\ collisions at $\sqrtsNN =
5.02$~\TeV\ over a wide kinematic range. The data were collected with
the ATLAS detector at the LHC and corresponds to $27.8$~nb$^{-1}$ of
integrated luminosity. The centrality of \pPb\ collisions was
characterised using the total transverse energy measured in the
forward calorimeter on the Pb-going side covering the interval $-4.9 <
\eta < -3.2$. The average number of nucleon--nucleon collisions and
the mean nuclear thickness factor were evaluated for each centrality
interval using a Glauber Monte Carlo analysis.

Results are presented for the nuclear modification factor \RpPb\ with
respect to a measurement of the inclusive jet cross-section in
$\sqrt{s}=2.76$~\TeV\ \pp\ collisions corresponding to $4.0$~pb$^{-1}$
of integrated luminosity. The \pp\ cross-section was \xT-interpolated
to $5.02$~\TeV\ using previous ATLAS measurements of inclusive jet
production at $2.76$ and $7$~\TeV. Results are also shown for the
central-to-peripheral ratio \Rcp. The centrality-inclusive
\RpPb\ results for 0--90\% collisions indicate only a modest
enhancement over the geometric expectation. This enhancement has a
weak \pT\ and rapidity dependence and is generally consistent with
predictions from the modification of the parton distribution functions
in the nucleus, which is small in the kinematic region probed by this
measurement.

The results of the \RCP\ measurement indicate a strong
centrality-dependent reduction in the yield of jets in central
collisions relative to that in peripheral collisions, after accounting
for the effects of the collision geometries. In addition, the
reduction becomes more pronounced with increasing jet \pT\ and at more
forward (downstream proton) rapidities. These two results are
reconciled by the centrality-dependent \RpPb\ results, which show a
suppression in central collisions and enhancement in peripheral
collisions, a pattern which is systematic in \pT\ and $y^*$.

The \RCP\ and \RpPb\ measurements at forward rapidities are also
reported as a function of $\pT\times\cosh(\left<y^*\right>)$, the
approximate total jet energy.  When plotted this way, the results from
different rapidity intervals follow a similar trend. This suggests
that the mechanism responsible for the observed effects may depend
only on the total jet energy or, more generally, on the underlying
parton--parton kinematics such as the fractional longitudinal momentum
of the parton originating in the proton.

If the relationship between the centrality intervals and proton--lead
collision impact parameter determined by the geometric models is
correct, these results imply large, impact parameter-dependent changes
in the number of partons available for hard scattering. However, they
may also be the result of a correlation between the kinematics of the
scattering and the soft interactions resulting in particle production
at backward (Pb-going)
rapidities~\cite{Alvioli:2014sba,Coleman-Smith:2013rla}.

Recently, the effects observed here have been hypothesized as arising
from a suppression of the soft particle multiplicity in collisions
producing high energy jets~\cite{Bzdak:2014rca}. Independently, it has
also been argued that proton configurations containing a large-$x$
parton interact with nucleons in the nucleus with a reduced
cross-section, resulting in the observed
modifications~\cite{Alvioli:2014eda}. In any case the presence of such
correlations would challenge the usual factorisation-based framework
for describing hard scattering processes in collisions involving
nuclei.

\section*{Acknowledgements}

We thank CERN for the very successful operation of the LHC, as well as the
support staff from our institutions without whom ATLAS could not be
operated efficiently.

We acknowledge the support of ANPCyT, Argentina; YerPhI, Armenia; ARC,
Australia; BMWFW and FWF, Austria; ANAS, Azerbaijan; SSTC, Belarus; CNPq and FAPESP,
Brazil; NSERC, NRC and CFI, Canada; CERN; CONICYT, Chile; CAS, MOST and NSFC,
China; COLCIENCIAS, Colombia; MSMT CR, MPO CR and VSC CR, Czech Republic;
DNRF, DNSRC and Lundbeck Foundation, Denmark; EPLANET, ERC and NSRF, European Union;
IN2P3-CNRS, CEA-DSM/IRFU, France; GNSF, Georgia; BMBF, DFG, HGF, MPG and AvH
Foundation, Germany; GSRT and NSRF, Greece; ISF, MINERVA, GIF, I-CORE and Benoziyo Center,
Israel; INFN, Italy; MEXT and JSPS, Japan; CNRST, Morocco; FOM and NWO,
Netherlands; BRF and RCN, Norway; MNiSW and NCN, Poland; GRICES and FCT, Portugal; MNE/IFA, Romania; MES of Russia and ROSATOM, Russian Federation; JINR; MSTD,
Serbia; MSSR, Slovakia; ARRS and MIZ\v{S}, Slovenia; DST/NRF, South Africa;
MINECO, Spain; SRC and Wallenberg Foundation, Sweden; SER, SNSF and Cantons of
Bern and Geneva, Switzerland; NSC, Taiwan; TAEK, Turkey; STFC, the Royal
Society and Leverhulme Trust, United Kingdom; DOE and NSF, United States of
America.

The crucial computing support from all WLCG partners is acknowledged
gratefully, in particular from CERN and the ATLAS Tier-1 facilities at
TRIUMF (Canada), NDGF (Denmark, Norway, Sweden), CC-IN2P3 (France),
KIT/GridKA (Germany), INFN-CNAF (Italy), NL-T1 (Netherlands), PIC (Spain),
ASGC (Taiwan), RAL (UK) and BNL (USA) and in the Tier-2 facilities
worldwide.

\bibliography{pPbJetPaper}
\bibliographystyle{elsarticle-num}

\onecolumn
\clearpage 
\begin{flushleft}
{\Large The ATLAS Collaboration}

\bigskip

G.~Aad$^{\rm 84}$,
B.~Abbott$^{\rm 112}$,
J.~Abdallah$^{\rm 152}$,
S.~Abdel~Khalek$^{\rm 116}$,
O.~Abdinov$^{\rm 11}$,
R.~Aben$^{\rm 106}$,
B.~Abi$^{\rm 113}$,
M.~Abolins$^{\rm 89}$,
O.S.~AbouZeid$^{\rm 159}$,
H.~Abramowicz$^{\rm 154}$
% ...
\bigskip
\\
% ...
$^{11}$ Institute of Physics, Azerbaijan Academy of Sciences, Baku, Azerbaijan\\
% ...
$^{84}$ CPPM, Aix-Marseille Universit{\'e} and CNRS/IN2P3, Marseille, France\\
% ...
$^{89}$ Department of Physics and Astronomy, Michigan State University, East Lansing MI, United States of America\\
% ...
$^{106}$ Nikhef National Institute for Subatomic Physics and University of Amsterdam, Amsterdam, Netherlands\\
% ...
$^{112}$ Homer L. Dodge Department of Physics and Astronomy, University of Oklahoma, Norman OK, United States of America\\
$^{113}$ Department of Physics, Oklahoma State University, Stillwater OK, United States of America\\
% ...
$^{116}$ LAL, Universit{\'e} Paris-Sud and CNRS/IN2P3, Orsay, France\\
% ...
$^{152}$ Institute of Physics, Academia Sinica, Taipei, Taiwan\\
% ...
$^{154}$ Raymond and Beverly Sackler School of Physics and Astronomy, Tel Aviv University, Tel Aviv, Israel\\
% ...
$^{159}$ Department of Physics, University of Toronto, Toronto ON, Canada\\
% ...
\end{flushleft}


\end{document}